\documentclass[pdflatex,sn-mathphys-num]{sn-jnl}

\usepackage{graphicx}%
\usepackage{multirow}%
\usepackage{amsmath,amssymb,amsfonts}%
\usepackage{amsthm}%
\usepackage{mathrsfs}%
\usepackage[title]{appendix}%
\usepackage{xcolor}%
\usepackage{textcomp}%
\usepackage{manyfoot}%
\usepackage{booktabs}%
\usepackage{algorithm}%
\usepackage{algorithmicx}%
\usepackage{algpseudocode}%
\usepackage{listings}%
\usepackage{float}
\usepackage{enumitem}

\begin{document}

\title[Article Title]{Boosting weather forecast via generative superensemble}


\author[1]{\fnm{Congyi} \sur{Nai}}\email{supernc226@gmail.com}

\author*[1]{\fnm{Xi} \sur{Chen}}\email{chenxi@lasg.iap.ac.cn}

\author[3,4]{\fnm{Shangshang} \sur{Yang}}

\author[5]{\fnm{Yuan} \sur{Liang}}

\author[2]{\fnm{Ziniu} \sur{Xiao}}

\author*[1]{\fnm{Baoxiang} \sur{Pan}}\email{panbaoxiang@lasg.iap.ac.cn}

\affil*[1]{\orgdiv{Key Laboratory of Earth System Numerical Modeling and Application},\orgdiv{Institute of Atmospheric Physics}, \orgname{Chinese Academy of Science}, \city{Beijing}, \country{China}}

\affil[2]{\orgdiv{Institute of Atmospheric Physics}, \orgname{Chinese Academy of Science}, \city{Beijing}, \country{China}}

\affil[3]{\orgdiv{Earth System Modelling}, \orgdiv{School of Engineering and Design}, \orgname{Technical University of Munich}, \city{Munich}, \country{Germany}}

\affil[4]{\orgdiv{Complexity Science},\orgname{Potsdam Institute for Climate Impact Research}, \city{Potsdam}, \country{Germany}}

\affil[5]{\orgname{TianJi Weather Science and Technology Company}, \city{Beijing}, \country{China}}




\abstract{
Accurate weather forecasting is essential for socioeconomic activities. While data-driven forecasting demonstrates superior predictive capabilities over traditional Numerical Weather Prediction (NWP) with reduced computational demands, its deterministic nature and limited advantages over physics-based ensemble predictions restrict operational applications. We introduce the generative ensemble prediction system (GenEPS) framework to address these limitations by randomizing and mitigating both random errors and systematic biases. GenEPS provides a plug-and-play ensemble forecasting capability for deterministic models to eliminate random errors, while incorporating cross-model integration for cross-model ensembles to address systematic biases. The framework culminates in a super-ensemble approach utilizing all available data-driven models to further minimize systematic biases. GenEPS achieves an Anomaly Correlation Coefficient (ACC) of 0.679 for 500hPa geopotential (Z500), exceeding the ECMWF Ensemble Prediction System's (ENS) ACC of 0.646. Integration of the ECMWF ensemble mean further improves the ACC to 0.683. The framework also enhances extreme event representation and produces energy spectra more consistent with ERA5 reanalysis. GenEPS establishes a new paradigm in ensemble forecasting by enabling the integration of multiple data-driven models into a high-performing super-ensemble system.}




\maketitle

\section{Introduction}\label{sec1}
Weather forecasting plays a pivotal role across socioeconomic sectors, with numerical weather prediction (NWP) serving as its foundation for seven decades \cite{asr-10-65-2013, magnusson2013factors,bauer2015quiet}. 
However, the emergence of data-driven approaches is now challenging this long-standing paradigm \cite{bi2023accurate,chen2023fengwu,chen2023fuxi,lam2023learning}. 
These models leverage extensive climate reanalysis data, high-capacity function approximators, and intensive optimization, so as to identify analogous patterns in weather trajectories. These learned patterns enable direct temporal \textit{jumps} in prediction, circumventing step-by-step resolving of geophysical fluid dynamics equations, achieving superior computational efficiency over traditional NWP approaches \cite{wolfram2003new,brunton2020machine}.

Despite their promise, data-driven weather forecasts face crucial challenges that currently limit their competitiveness against state-of-the-art numerical ensemble forecasting systems.
These models typically produce over-smoothed meteorological fields, leading to underestimation of extreme events and poor resolving of fine-scale features \cite{rasp2024weatherbench}. Additionally, they often lack robust uncertainty quantification — a crucial component for operational forecasting decision-making \cite{palmer2002economic}. These limitations become more pronounced in auto-regressive forecasting, where errors accumulate over extended prediction horizons. 

These limitations can be traced to a common root cause: inadequate uncertainty representation during model training and application. While most data-driven forecasting approaches optimize for minimal aggregate prediction error across cases, variables, locations, and lead times \cite{chen2023fengwu,chen2023fuxi,lam2023learning}, this aggregation obscures three distinct uncertainty sources \cite{bauer2015quiet,buizza2019introduction}: initial state estimation uncertainty, arising from incomplete observations and assimilation deficiencies; model stochasticity, resulting from insufficient resolution of geophysical fluid dynamics and simplified geophysical processes; and model bias, reflecting systematic errors in parameterizing unresolved processes. The first two represent irreducible random uncertainties, best characterized through probability distributions; while the third constitutes a systematic error, specific to individual model configurations, and could potentially be eliminated.

Failure to distinguish and account for these different uncertainty sources through a forecasting life-cycle leads to their amplification and nonlinear propagation via complex feedbacks \cite{bauer2015quiet}. This results in state-dependent uncertainty growth patterns in high-dimensional space, 
ultimately rending uncertainty quantification and error correction intractable.

Superensemble forecasting \cite{krishnamurti1999improved,hagedorn2005rationale} has emerged as a promising solution, characterizing uncertainty evolution by comprehensive sampling across three dimensions: initial states, stochastic process representations, and model formulations. Initial state sampling occurs at forecast initiation, while stochastic and model formulation sampling continue throughout the forecasting period, with their effects systematically integrated into the final uncertainty quantification.

Numerical prediction faces crucial constraints in implementing superensemble approaches. The computational cost of large ensemble simulations remains prohibitive. Moreover, to inject model noise - via stochastic parameterizations - requires a delicate balance: these schemes must accurately represent sub-grid tendency, while introducing appropriate stochasticity levels. Additionally, model-specific state representations prevent direct transfer of intermediate states among models, confining forecast trajectories within individual frameworks and limiting comprehensive uncertainty sampling.

Recent data-driven approaches have attempted to address these limitations by leveraging advances in deep generative modeling \cite{pan2022improving,price2023gencast,zhong2024fuxi}. These methods aim to construct conditional probability distribution of future weather states, rather than deterministic forecast. However, several fundamental challenges persist.   
While a conditional probability estimator samples stochastic uncertainty for an individual forecasting model, it requires extra treatment on initial state uncertainty, and misses the opportunity to leverage multi-model ensembles' potential for systematic bias cancellation through diverse model formulations.
Moreover, training such conditional probability distribution models requires substantially greater computational resources compared to deterministic forecasting models, yet lacks reusability: any modification in input variable choice, model architectural or hyperparameter setup, requires a complete retraining. 


We present generative ensemble prediction system (GenEPS), a plug-and-play solution to the above-mentioned challenges.  
GenEPS learns an unconditional generative model of high-dimensional atmospheric states. Using this states distribution as an informative prior, we develop a simple approach to flexibly combine it with initial weather state guess and a pool of forecasting models, so as to generate superensemble weather forecasts, sampling all categories of uncertainty sources.

This yields state-of-the-art deterministic and probabilistic weather forecasting skills. 
The efficacy of GenEPS is validated through diverse experiments. Large-scale predictive skill was assessed using 500 hPa geopotential anomaly correlation coefficients (ACC). Real-world predictive performance was evaluated against in situ observational data. Extreme event forecasting capabilities were examined through case studies of heatwaves and tropical cyclones. 
\section{GenEPS: generative ensemble prediction system}\label{sec2}

\begin{figure}[H]
\centering
\centerline{\includegraphics[width=\linewidth]{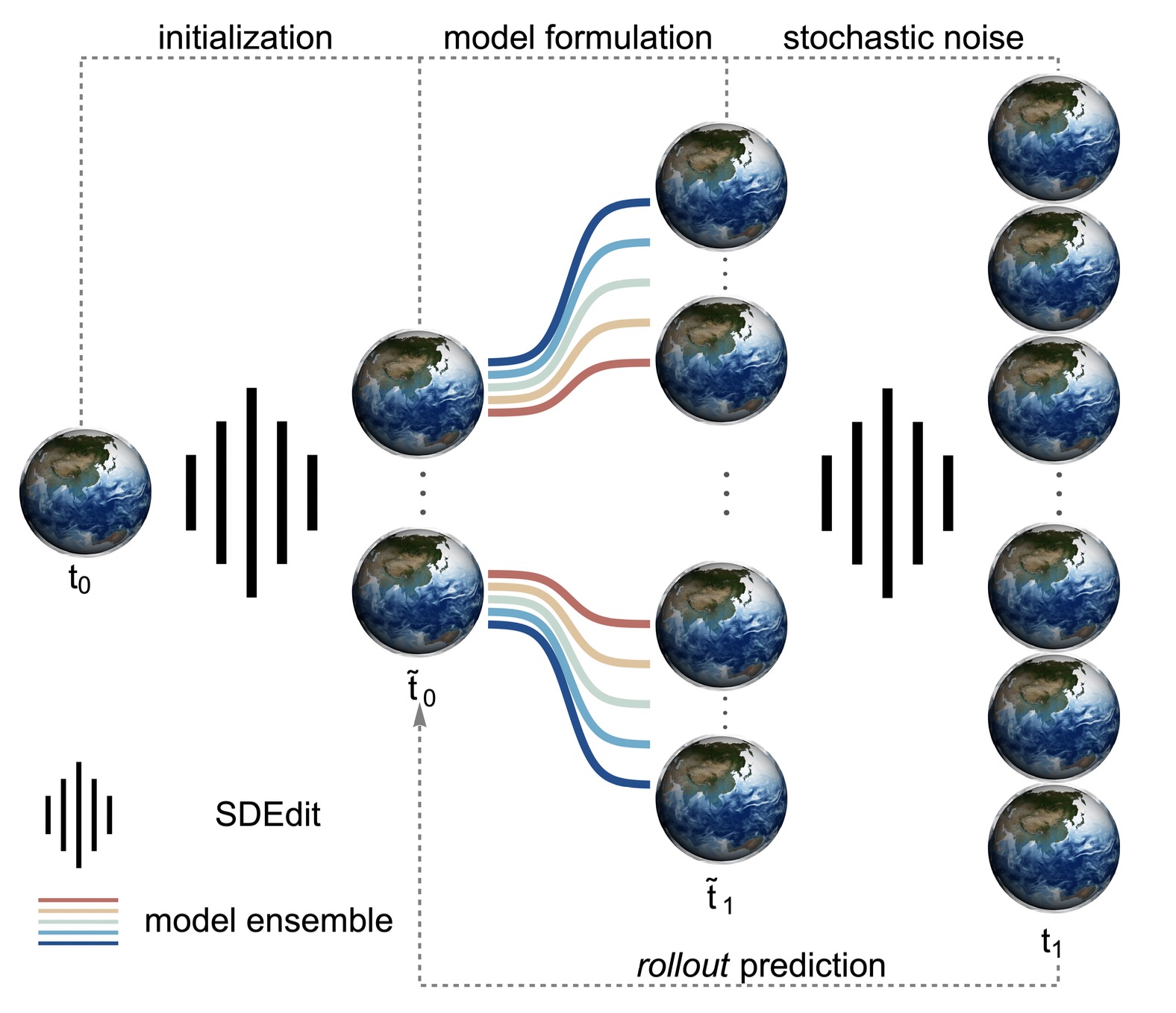}}
\caption{Schematic illustration of the GenEPS workflow, demonstrating three key sources of uncertainty in weather forecasting. Initial condition uncertainty is addressed through SDEdit-based perturbations at $t_0$, generating perturbed states $\tilde{t}_0$. Model formulation uncertainty is sampled through multiple forecast models (indicated by different colored paths) evolving from each perturbed state to intermediate states $\tilde{t}_1$. Stochastic uncertainty is then incorporated at prediction time $t_1$, resulting in the final ensemble forecast. This multi-stage approach enables comprehensive sampling of the forecast uncertainty space.}\label{fig10086}
\end{figure}

The atmospheric state at any given time can be viewed as a high-dimensional point in phase space, with its degrees of freedom determined by spatial resolution and occurrence frequency. We employ probabilistic diffusion models \cite{sohl2015deep,ho2020denoising,songscore} to characterize the high dimensional atmospheric state probability density function, aligning with historical reanalysis data. The learned prior is subsequently utilized for posterior inference\cite{meng2021sdedit}, enabling atmospheric state perturbations that better align with historical data. This approach serves three distinct purposes: 1) To sample initial condition uncertainty by generating ensemble perturbations that are consistent with the learned atmospheric state distribution; 2) To sample model stochastic uncertainty by modifying intermediate states during model integration to better align with the objective atmospheric state distribution; 3) To “decouple” the model-state dependencies, enabling seamless integration across different models from arbitrary restart states. This flexibility in model switching allows finite-member models to sample a broader spectrum of model behaviors through interactive integrations, thereby enhancing the representation of model formulation uncertainty space and improving the characterization of the true probability density function of atmospheric states. The GenEPS represents a novel paradigm in probabilistic multi-model ensemble forecasting. The complete workflow of GenEPS, illustrating the sampling of initial condition, model formulation, and stochastic uncertainties, is depicted in Figure\ref{fig10086}.

\section{Multi-Source Uncertainty Sampling}\label{sec2}

Ensemble forecasting requires comprehensively sampling the high-dimensional probability distribution function (PDF) of forecast states by accounting for multiple sources of uncertainty. When this uncertainty estimation achieves sufficient accuracy, ensemble averaging can effectively cancel random errors. However, most current deep learning models either remain purely deterministic without uncertainty consideration, or only partially account for model stochastic uncertainty, leading to inadequate characterization of the high-dimensional uncertainty space.

\begin{figure}[h]
\centerline{\includegraphics[width=\linewidth]{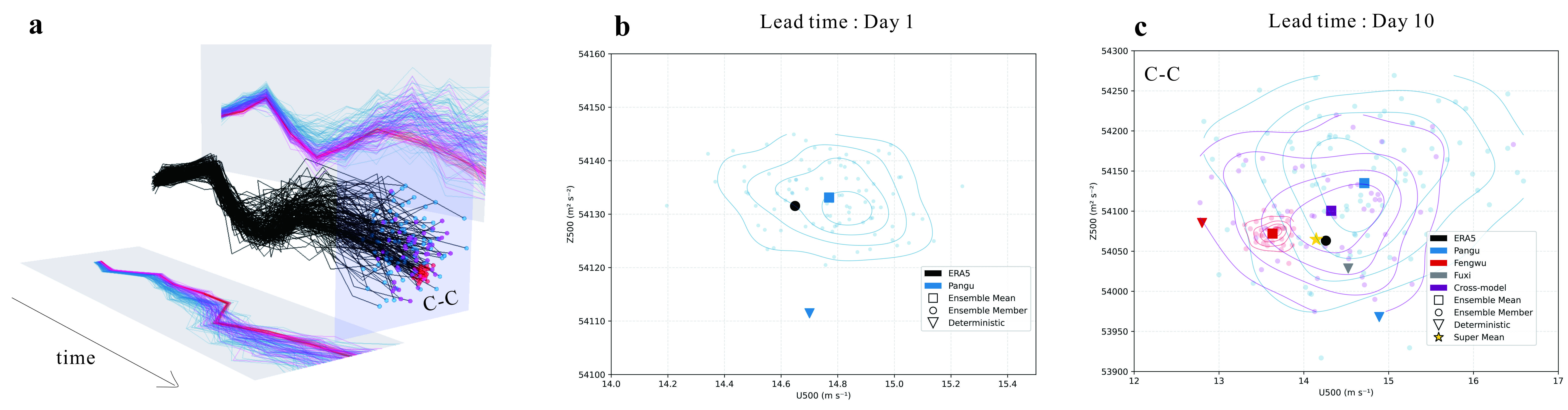}}
\caption{(a) Low-dimensional visualization of atmospheric state evolution with interannual variability removed through temporal averaging, projected onto a single spatial point of Z500 and U500 (selected from the full state space of UVTQZ at 13 pressure levels and surface variables T2m, SLP, U10m, V10m at 0.25° resolution); (b) Cross-section of atmospheric evolution trajectories at day 1, with points representing ensemble member states at this forecast lead time.; (c) Cross-section of ensemble forecast states at day 10, showing the distribution of different ensemble types (single-model ensembles in red and blue, cross-model ensemble in purple) relative to ERA5 verification (black triangle). Squares denote ensemble means, and the gold pentagram represents the GenEPS grand mean.}\label{fig1}
\end{figure}

Consider a high-dimensional atmospheric state comprising upper-air U,V,T,Q,Z and surface (T2m,slp,10mu,10mv) fields at 0.25-degree resolution, encompassing (13×5+4)×721×1440=71,638,560 degrees of freedom. To visualize the high-dimensional space, Figure.\ref{fig1} projects the atmospheric state onto a phase space defined by 500-hPa geopotential and zonal wind at a randomly selected grid point, depicting the GenEPS ensemble trajectories (Figure.\ref{fig1}a). 


As illustrated in Figure.\ref{fig1}b, a cross-section of selected day-1 forecast trajectories compares deterministic Pangu forecasts against forecasts with initial condition perturbations generated through the Generative State Matching (GSM, detailed in \ref{subsec1}). GSM produces initial perturbations that demonstrate physical consistency - the resulting ensemble members show appropriate spread around the verification state without excessive dispersion, while effectively encompassing the ERA5 verification state and achieving reduced ensemble mean error.


Furthermore, GSM can simultaneously perturb both initial and intermediate integration states, thereby characterizing uncertainties from both initial conditions and model stochasticity. This versatility enables GSM to serve as a plug-and-play ensemble generation framework for any deterministic forecast model. As demonstrated in Figure.\ref{fig1}c, the GSM-generated ensemble members for both Fengwu and Pangu models (shown as red and blue circles) effectively encompass the ERA5 verification state. The ensemble means (represented by red and blue squares) show reduced errors compared to their respective deterministic forecasts, validating the effectiveness of this uncertainty estimation approach. After ensemble averaging successfully cancels random errors, the remaining deviations in the ensemble mean states primarily reflect systematic biases.


The remaining systematic biases in ensemble mean states underscore the importance of characterizing model-specific uncertainty. While simply combining multiple model ensembles can partially represent this uncertainty, the limited number of distinct model behaviors leads to insufficient characterization of the model-specific uncertainty space. A key technical challenge in developing cross-model methods is that intermediate forecast states are not directly transferable between models due to inconsistent error accumulation patterns.  GSM overcomes this limitation by enhancing the consistency between model intermediate states and objective atmospheric distributions, effectively decoupling atmospheric states from model-specific systematic biases. This decoupling enables random model switching during integration, allowing for more comprehensive sampling of the model-specific uncertainty space with a limited number of models. The resulting cross-model ensemble (purple circles in Figure.\ref{fig1}c) produces novel model behaviors distinct from single-model ensembles (red or blue circles),  This improved uncertainty representation is reflected in the cross-model ensemble mean (purple square) showing smaller deviation from ERA5 compared to either the Pangu or Fengwu ensemble means.

GSM and cross-model ensemble techniques successfully isolate and eliminate random errors while providing preliminary sampling of model-specific systematic uncertainties, with ensemble averaging demonstrating effective error reduction. However, no model exists without systematic biases, and increased ensemble sizes facilitate more comprehensive characterization of the uncertainty space. The integration of all ensemble members (Pangu Ensemble, Fengwu Ensemble, and cross-model ensemble) enables a comprehensive representation of forecast state uncertainties. This multi-faceted sampling process across three distinct uncertainty spaces forms the foundation of the GenEPS. The effectiveness of this comprehensive uncertainty characterization is demonstrated by the ensemble grand mean (denoted by gold pentagram in Figure.\ref{fig1}c), which is derived from aggregating ensemble means and deterministic forecasts and exhibits minimal deviation from ERA5. 

In essence, GenEPS integrates diverse uncertainty samplings, suggesting potential for enhanced forecast skill through improved uncertainty characterization. While the results presented here demonstrate improved uncertainty representation in these selected low-dimensional projections, comprehensive validation requires evaluation across the full high-dimensional state space using conventional verification metrics, which will be addressed in subsequent sections.

\section{Ensemble forecasting skill}\label{sec2}
 
\begin{figure}[h]
\centerline{\includegraphics[width=\linewidth]{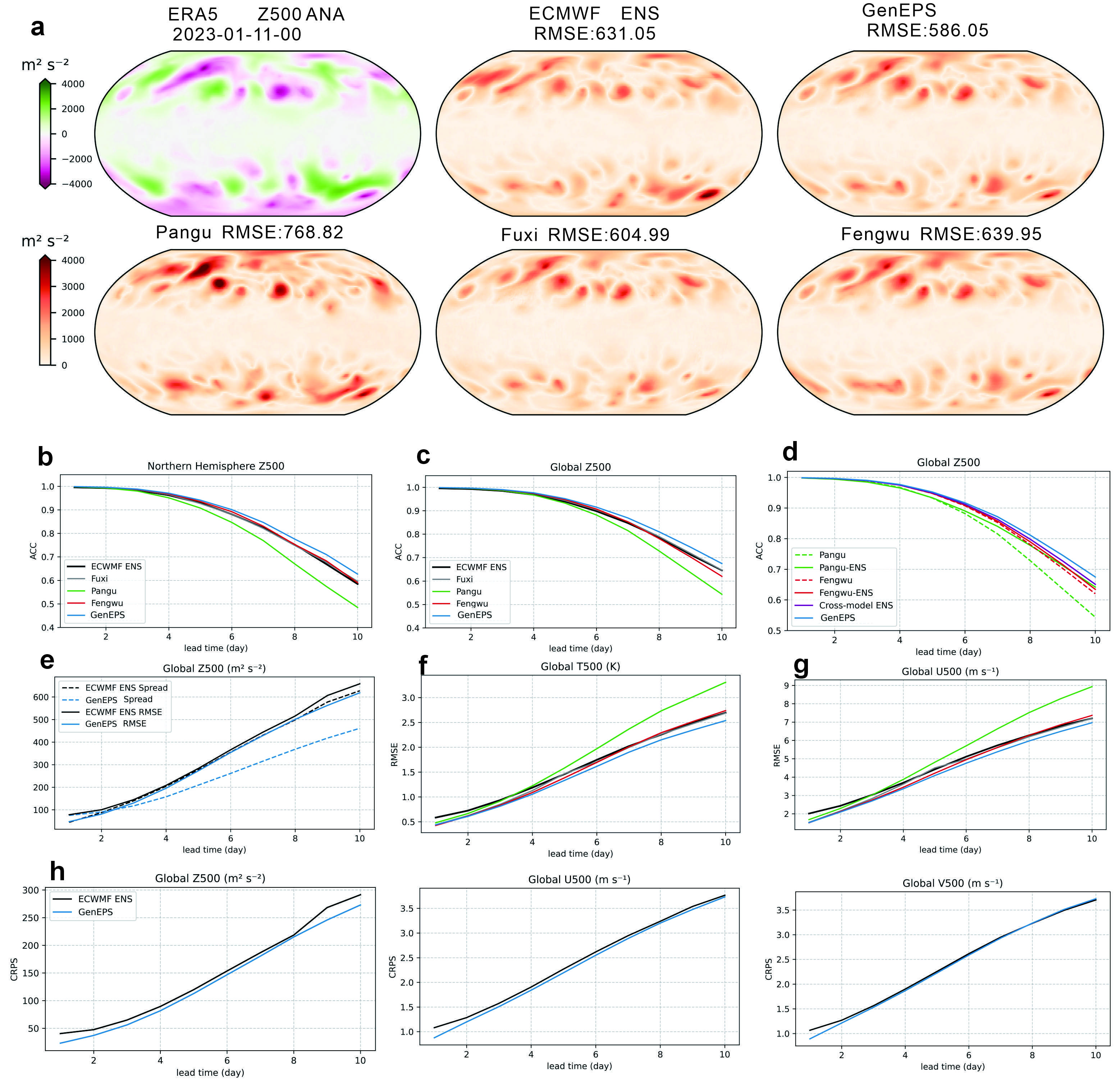}}
\caption{(a) 500hPa geopotential anomaly for 2023-01-11-00: ERA5 (truth, top-left) compared with day-10 absolute errors from ECMWF Ensemble (50-member mean), GenEPS (240-member mean), Pangu, Fuxi, and Fengwu, with corresponding RMSE values; (b) Northern Hemisphere Z500 ACC; (c) Global Z500 ACC; (d) Skill improvement through single-model ensembles and cross-model ensembles; (e) RMSE-Spread comparison between ECMWF ENS (50 members) and GenEPS (240 members); (f,g) T500 and U500 RMSE for different models; (h) CRPS comparison between GenEPS and ECMWF ENS for Z500, U500, and V500.}\label{fig2}
\end{figure}

 This section evaluates both deterministic and probabilistic skills of the GenEPS against various benchmarks at 1.5-degree resolution. Figure.\ref{fig2}a illustrates the spatial distribution of absolute errors through a case study of 10-day Z500 anomaly forecasts. In this case, the GenEPS demonstrates superior performance with a 7 percent lower Root Mean Square Error (RMSE) compared to the ECMWF ensemble mean, achieving the highest accuracy among all evaluated models.

Figure.\ref{fig2}b and Figure.\ref{fig2}c compare Z500 Anomaly Correlation Coefficients (ACC) across different models for Northern Hemisphere and global regions. GenEPS demonstrates superior forecast skill, achieving a global Z500 ACC of 0.679 at day 10, exceeding ECMWF ENS by 5.1 percent. Including ECMWF mean as an independent member further improves the 10-day ACC to 0.683. Spectral analysis (supporting Figure 1) reveals that both FENGWU and FUXI exhibit characteristic ensemble-mean behavior, with energy spectra matching ECMWF ENS mean. FUXI shows notable spectral discontinuity in Z500 and U500 fields between days 5-6, corresponding to its short-to-medium range model transition. This ensemble-like behavior explains their comparable forecast skill to ECMWF ENS.

Figure.\ref{fig2}d illustrates the reduction of random errors and systematic biases in forecast skill. The GenEPS framework-based PanguENS and FuxiENS (solid green and red lines, respectively) demonstrate superior skill compared to their deterministic counterparts Pangu and Fengwu (dashed green and red lines) through ensemble averaging, which effectively reduces random errors. Cross-model Ens, derived by sampling from Pangu and Fengwu at each forecast step (solid purple line), achieves preliminary mitigation of systematic biases, yielding enhanced ACC. The final skill (solid blue line) represents further improvement through multi-model averaging, indicating additional systematic bias reduction.

Figure.\ref{fig2}e illustrates Z500 RMSE and spread characteristics for GenEPS and ECMWF ENS.While GenEPS demonstrates superior RMSE performance, it exhibits smaller spread than ECMWF ENS, which employs maximum error growth perturbations to achieve larger spread with limited ensemble members. However, ECMWF ENS's approach typically results in excessive initial perturbations, significantly degrading forecast skill in the first three days, as evidenced by higher RMSE in Figure.\ref{fig2}e, f, and g. GenEPS achieves sufficient spread through iterative perturbations while minimizing initial state disruption, maintaining short-range forecast skill and achieving superior performance with smaller spread, suggesting potential for further improvement through increased spread.

Figure.\ref{fig2}h presents the Continuous Ranked Probability Score (CRPS) comparison for Z500, U500, and V500 between GenEPS and ECMWF ENS. GenEPS demonstrates superior probabilistic forecast skill, particularly evident in the first three days of the forecast period.



\section{Site-Specific Observational Evaluation}\label{sec3}

While reanalysis products offer valuable insights into atmospheric states, they cannot perfectly reflect true atmospheric conditions. To assess the "real" forecasting skill of various prediction products,  we utilized ground-based observations from over 2,168 stations across China to validate 10-day 10-meter wind speed forecasts.

\begin{figure}[H]
\centering
\centerline{\includegraphics[width=\linewidth]{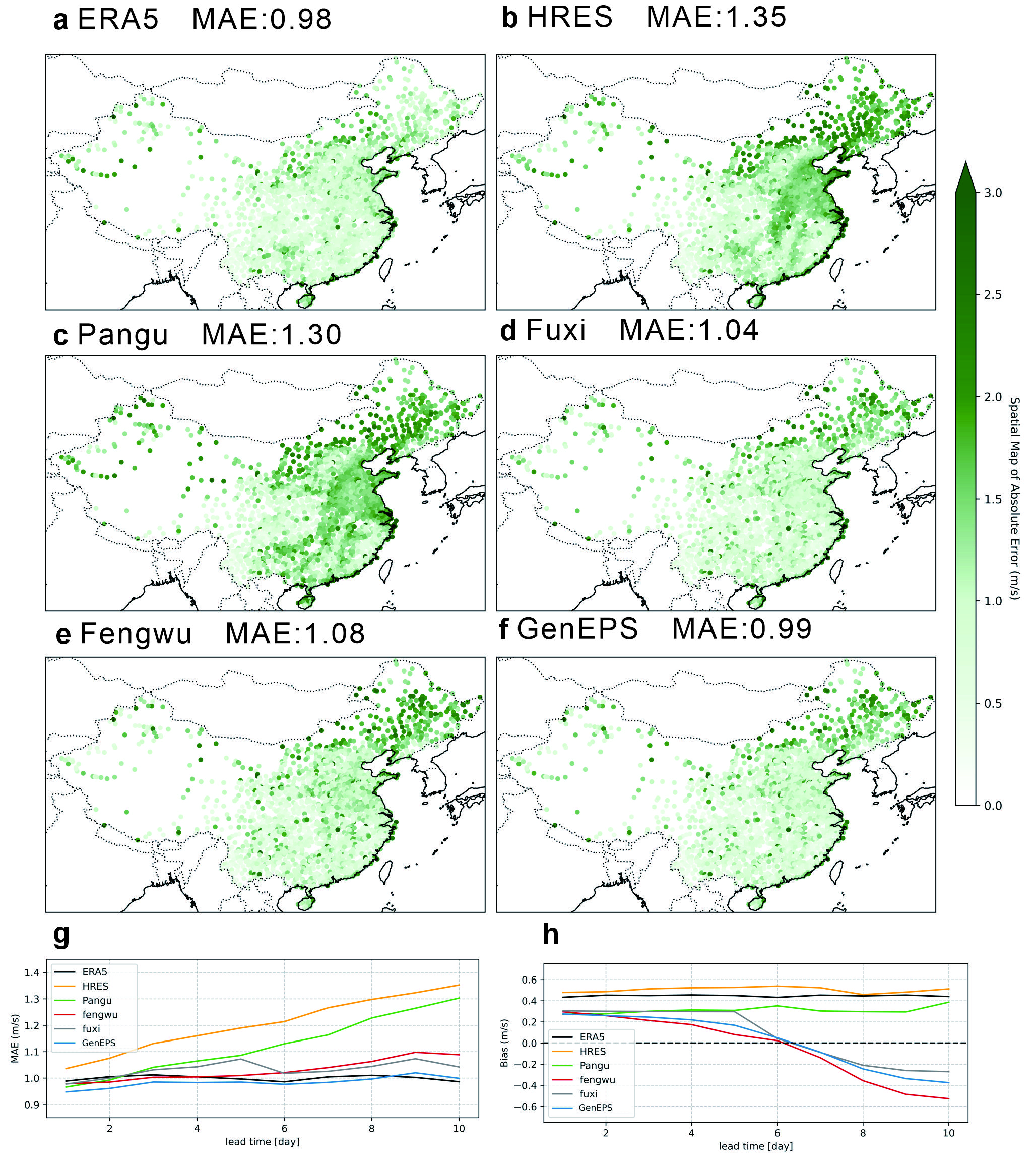}}
\caption{Evaluation of 10-meter wind speed predictions using Chinese surface station data.(a-d) Spatial distribution of Mean Absolute Error (MAE) for 10-meter wind speed forecasts compared to observed values. Results are shown for (a) ERA5 reanalysis, (b) IFS forecast, (c-e) Pangu, Fuxi and Fengwu forecast. (f) GenEPS mean (g) Bias in 10-meter wind speed forecasts. (h) Root Mean Square Error of 10-meter wind speed forecasts.)}\label{fig3}
\end{figure}

Figure.\ref{fig3} illustrates spatial absolute errors in day-10 forecasts across models. ERA5 reanalysis shows minimal errors but maintains systematic bias relative to site observations. Compared to Fuxi and Fengwu, which exhibit ensemble-like behavior, Pangu and HRES demonstrate notable errors over North China. GenEPS achieves the lowest error (0.99 m/s) among all forecast models, approaching ERA5's error (0.98 m/s). This similarity likely stems from all data-driven models being trained on ERA5, thus inheriting comparable systematic biases relative to "true" observations.

The temporal evolution of forecast metrics, illustrated in panels g and h, provides further insights into model performance over the 10-day forecast period. While HRES and Pangu show increasing MAE over time, the GenEPS maintains relatively stable performance comparable to ERA5. While, most ensemble mean forecasting including GenEPS, tend towards a negative bias by day 10. This consistent underprediction of wind speeds in the extended range may be attributed to the ensemble averaging process. The smoothing effect inherent in ensemble methods can lead to a reduction in predicted extreme values, potentially resulting in an overall negative bias for wind speed forecasts.

These findings highlight the efficacy of the GenEPS. By integrating multiple models, the GenEPS achieves performance on par with the ERA5 reanalysis in terms of MAE. This is particularly significant given that ERA5, as a reanalysis product incorporating observational data, typically outperforms pure forecasts. The remarkable stability of the GenEPS in both RMSE and bias over the 10-day forecast period suggests its potential for reliable extended-range forecasting, likely due to its ability to balance out individual model biases.

\section{Extreme events}\label{sec4}
Data-driven models in weather forecasting are often criticized for their inadequate representation of extreme events. This section demonstrates how GenEPS addresses this challenge by leveraging the computational efficiency of machine learning to generate large ensemble memberships (utilizing Pangu as the backbone model due to its computational efficiency), enabling improved coverage of extreme weather events.

\subsection{Heat wave}

\begin{figure}[H]
\centering
\centerline{\includegraphics[width=\linewidth]{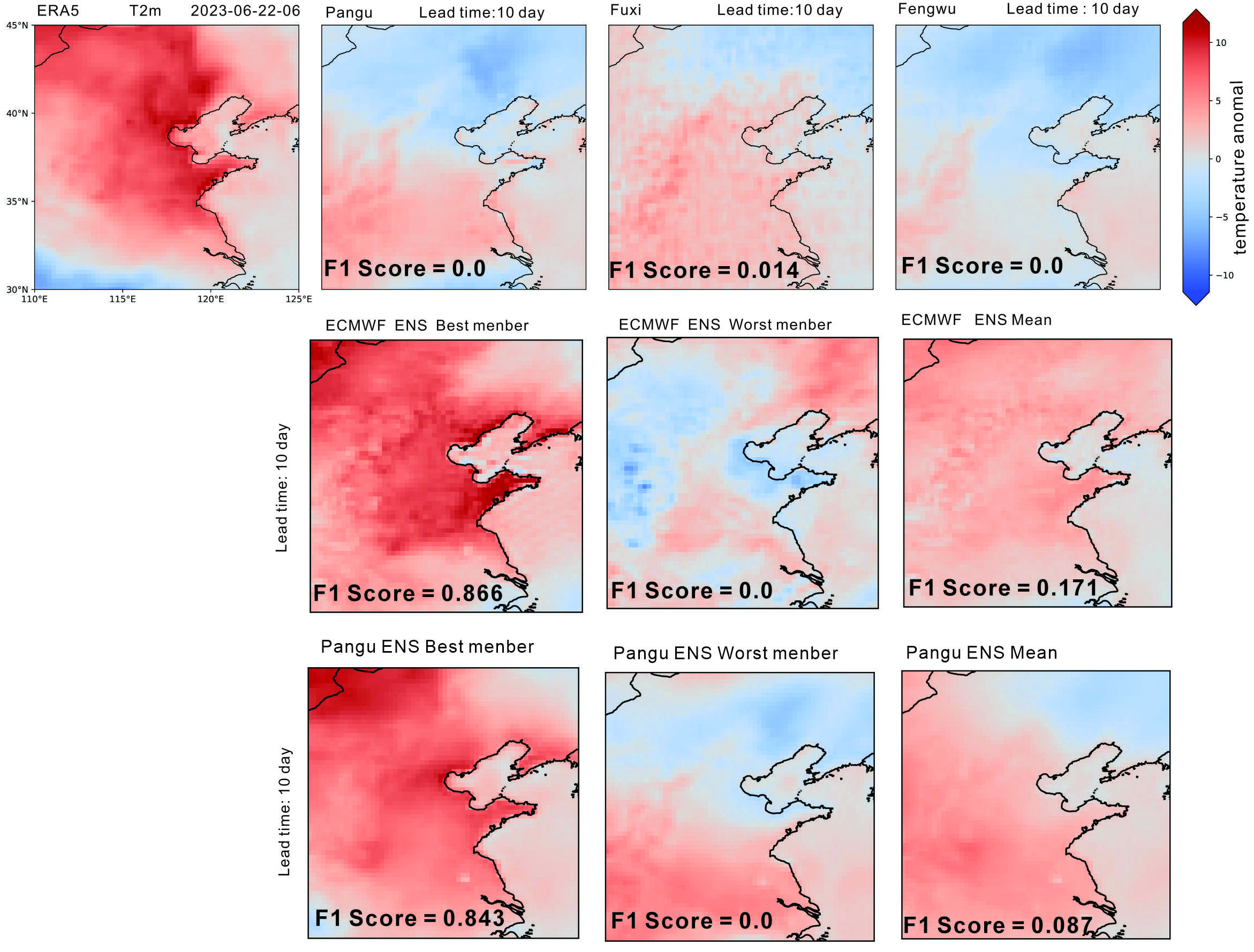}}
\caption{The top row displays the ERA5 reanalysis (leftmost) followed by three different predictions from Pangu, Fuxi, Fengwu. The middle and bottom rows show forecasts from ECMWF ENS and Pangu ENS models respectively, including best member, worst member, and mean predictions. F1 scores are provided for each prediction,  calculated using a threshold of $>$5°C temperature anomaly.}\label{fig4}
\end{figure}

Figure.\ref{fig4} presents a comparative analysis of various forecasting models in predicting the extreme heat wave that affected North China on June 22, 2023. The figure displays 2-meter temperature anomaly from ERA5 reanalysis, deterministic models (Pangu, Fuxi, Fengwu), and ensemble forecasts from both the ECMWF ENS and Pangu ENS, all for a 10-day forecast lead time.

The ERA5 reanalysis (top left panel) clearly depicts the intensity and spatial extent of the heat wave, serving as a reference for evaluating forecast skill. The deterministic models (Pangu, Fuxi, and Fengwu) demonstrate significant limitations in capturing this extreme event, as evidenced by their low F1 scores (0.0, 0.014, and 0.0 respectively). This underperformance highlights a critical challenge in predicting extreme weather phenomena using deterministic forecasts, particularly for extended lead times.

In contrast, the ensemble-based approaches show markedly improved capabilities in representing the potential for extreme conditions. The ECMWF ENS exhibits a wide range of outcomes, with its best member achieving a high F1 score of 0.866, closely approximating the observed heat wave pattern. Pangu-Ensemble demonstrates effective capability in capturing extreme events, significantly enhancing predictive skill compared to deterministic data-driven forecasts, with its best member achieving an F1 score of 0.843. The performance nearly matches that of the physics-based ECMWF system.

However, the Pangu ENS mean shows a lower skill compared to the ECMWF ENS mean, which highlights the ongoing advantages of physically-based models in representing extreme events. This discrepancy may be attributed to the limited training data (only three years ERA5) used in developing the prior atmospheric state distribution, which may not fully capture the characteristics of rare, extreme events like the North China heat wave of June 2023.

These findings highlight both the potential and current limitations of machine learning-based ensemble forecasting systems. While the GenEPS shows promise in generating diverse and skillful ensemble members, its performance in extreme event scenarios is constrained by the limited historical data used in training.

\subsection{Tropical cyclone}\label{subsec5}

Precise forecasting of tropical cyclones holds immense practical significance, with direct implications for public safety, disaster preparedness, and economic planning in affected regions. Recently, deterministic data-driven models have demonstrated their capacity to surpass the performance of traditional high-resolution (HRES) numerical weather prediction models in tropical cyclone track prediction. However, these data-driven models have been limited by their inability to provide uncertainty estimates and accurately predict extreme values associated with such intense weather systems. This subsection demonstrates the efficacy of the GenEPS in tropical cyclone forecasting, particularly for Typhoon Doksuri in 2023.

\begin{figure}[H]
\centering
\centerline{\includegraphics[width=\linewidth]{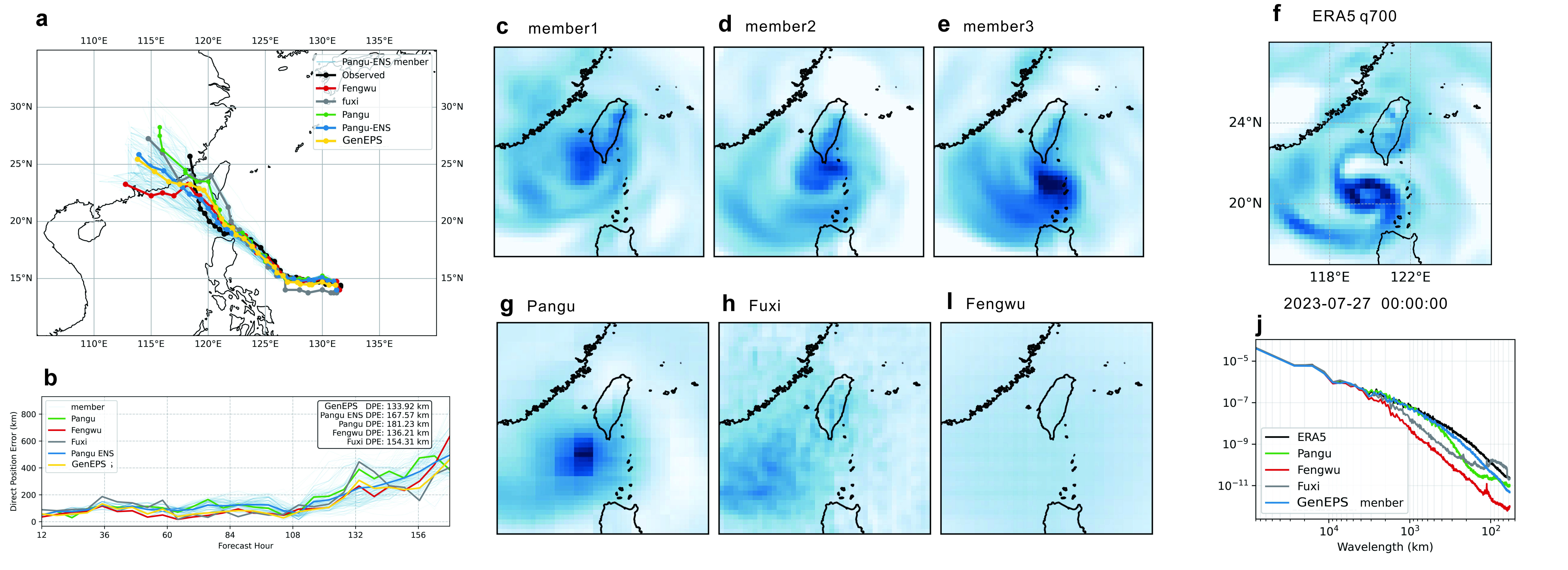}}
\caption{(a) Track forecasts of Typhoon "Doksuri" from different models; (b) Direct Position Error (DPE) comparison across models; (c-e) Day-7 forecasts of 700hPa specific humidity (q700) from individual Pangu-ENS members; (f) ERA5 verification of q700; (g-i) Day-7 q700 forecasts from deterministic models: Pangu, Fuxi, and Fengwu; (j) Spectral analysis of different fields.}\label{fig5}
\end{figure}

Figure.\ref{fig5}a displays forecast tracks from different models alongside the observed best track.  Pangu ENS (256 ensemble members) encompasses the best track, providing useful uncertainty estimation. Figure.\ref{fig5}b illustrates the temporal evolution of track errors, with Pangu-ENS demonstrating improved track prediction compared to deterministic Pangu. The GenEPS, averaging multiple model (Fuxi, Fengwu and Pangu-ENS) tracks, achieves the lowest mean position error (133km).

presents a comprehensive analysis of various forecasting models in predicting the track and structure of a tropical cyclone. The Pangu ENS, comprising 256 members generated using the Pangu model as a backbone, demonstrates superior performance with the lowest mean track error of 133 km. Panel (a) displays the forecasted tracks from different models alongside the observed best track, while panel (b) illustrates the evolution of track errors over time. The consistent outperformance of GenEPS highlights the benefits of its large member set in capturing the range of possible outcomes and providing valuable uncertainty information.

Figure.\ref{fig5}c-j analyze cyclone structure at day 7 (700hPa specific humidity). Deterministic forecasts from Fuxi and Fengwu (panels g,h) fail to reproduce cyclonic structure and lack extremes. While Pangu's forecast (panel i) maintains some intensity, it poorly represents typhoon structure. This deficiency is reflected in the energy spectra (panel j), showing significant deviations from ERA5, particularly at smaller scales, indicating substantial loss of fine-scale information. All three models exhibit grid discontinuities or checkerboard effects, artifacts from training with 4x4 linear patches on 0.25° resolution data, manifesting as spectral energy unphysical peak around 100km scales. These results highlight critical limitations of current deterministic data-driven models in representing extreme events. However, Pangu-ENS members under the GenEPS framework largely mitigate these issues, maintaining extremes at day 7 with improved cyclone structure representation and spectral characteristics more closely aligned with ERA5 across all scales.

\section{Discussion}\label{sec4}

Deterministic data-driven models have made medium-range weather forecasting more accessible, enabling the generation of forecasts comparable to IFS HRES performance with significantly reduced computational costs. This study further advances the accessibility of ensemble forecasting, enabling researchers to achieve medium-range ensemble predictions that surpass ECMWF ENS performance using relatively modest computational resources.


This study introduces a novel generative ensemble prediction system that employs probabilistic diffusion models to learn the prior distribution of high-dimensional atmospheric states from ERA5 reanalysis data. The learned prior is subsequently integrated with SDEdit techniques to perform posterior inference by modifying forecast fields or initial conditions. Sampling from both initial condition and model stochastic uncertainty spaces enables ensemble forecasting capabilities for deterministic data-driven models. The ensemble averaging process effectively cancels out random errors, leading to enhanced forecast skill.

Single-model ensemble averaging effectively eliminates most random errors, leaving systematic biases as the predominant form of remaining errors. Sampling in model formulation uncertainty space provides a way to address otherwise intractable model-specific systematic biases, which transforms model-specific systematic biases into random errors across multiple model behaviors. Through generative state matching, which decouples intermediate forecast states from model biases, forecasts can be continued using different models (Pangu and Fengwu in this work) during the forecasting process, providing a preliminary solution for sampling model-specific uncertainty to mitigate systematic biases. Further averaging across all models (Pangu, Fengwu and Fuxi in this work) and their ensemble means produces a superensemble, which further enhances forecast skill. Under this configuration, the ACC for 500 hPa geopotential at day 10 reaches 0.679, representing a 5.1 percent improvement compared to the ECMWF ENS, with a further increase to 0.683 when incorporating the ECMWF ENS.

Increasing both ensemble members and model diversity enhances the characterization of random error and systematic bias PDF, thereby strengthening the mitigation of random and systematic errors. GenEPS framework leverages machine learning computational efficiency to facilitate high-throughput multi-model ensemble generation, improving the representation of random error in state space. Moreover, GenEPS employs cross-model ensemble to generate diverse model behaviors from a limited number of models, enabling more comprehensive sampling of the model-specific systematic uncertainty space.

GenEPS has demonstrated remarkable skill in medium-range large-scale forecasting. However, several limitations persist. Primarily, the enhanced large-scale performance is achieved at the expense of small-scale features; the preservation of fine-scale information while maintaining large-scale forecast skill remains an unresolved challenge. Additionally, while GenEPS substantially improves the predictive capabilities of deterministic data-driven models for extreme events, their performance still falls short compared to physics-based models. This limitation likely stems from insufficient training data and computational constraints. Furthermore, modeling the high-dimensional state space at 0.25-degree resolution presents significant engineering challenges, necessitating compromises in both iteration counts and model parameterization due to computational and storage capacity limitations.

Future research directions should encompass two primary aspects: (1) implementation of physics-based models or machine learning downscaling techniques to reconstruct sub-grid scale information within improved large-scale forecasts, aiming to generate forecast products that simultaneously achieve high large-scale predictive skill and appropriate energy spectral characteristics; and (2) scaling up the current model through enhanced computational resources and expanded datasets to improve the representation and prediction of extreme events.

\section{Methods}\label{sec5}

\subsection{ERA5 training and evaluation data }\label{Appendix_sec_A1}
ERA5 represents the state-of-the-art global reanalysis dataset. We posit that ERA5 most accurately represents the objective atmospheric state PDF. The modeling target comprises 69 variables at 0.25-degree resolution (721×1440 latitude-longitude grid points). The variable set encompasses five upper-air atmospheric parameters across 13 pressure levels (50, 100, 150, 200, 250, 300, 400, 500, 600, 700, 850, 925, and 1000 hPa), supplemented by four surface variables. The upper-air variables include geopotential (Z), specific humidity (Q), temperature (T), zonal wind component (U), and meridional wind component (V). Surface parameters comprise 2-meter temperature (T2M), 10-meter zonal wind component (U10), 10-meter meridional wind component (V10), and mean sea-level pressure (MSL). Table 1 summarizes the modeled variables and input parameters (primarily providing spatial information).

The model training utilized data spanning 2018-2022, with 2023 data reserved for testing. To optimize ensemble size within computational constraints while preserving annual cyclicity, test cases were selected at 0000 UTC at five-day intervals throughout 2023 (i.e., 0000 UTC 1 January 2023, 0000 UTC 6 January 2023, and so forth), yielding 72 forecast initialization times.

\begin{table}[h]
\centering
\begin{tabular}{lccc}
\toprule
\textbf{Type} & \textbf{Variable name} & \textbf{Short name} & \textbf{Role} \\
\midrule
upper-air variables & Geopotential  & z  & output  \\
upper-air variables          & Specific humidity                   & q                   & output   \\
upper-air variables          & Temperature                         & t                   & output      \\
upper-air variables          & U component of wind                 & u                   & output       \\
upper-air variables          & V component of wind                 & v                   & output         \\
\midrule
surface variables               & 2 metre temperature                 & 2t                  & output      \\
surface variables               & 10 metre u wind component           & 10u                 & output       \\
surface variables               & 10 metre v wind component           & 10v                 & output      \\
surface variables               & Mean sea level pressure             & msl                 & output          \\
\midrule
geographical               & land-sea mask             & lsm                 & intput          \\
geographical               & topography             & tp                 & intput          \\
geographical               & latitude             & lat                 & intput          \\
geographical               & longitude             & lon                 & intput          \\
\bottomrule
\end{tabular}
\caption{A summary of all input and output variables. The "Role" column differentiates between target variables for modeling and auxiliary input parameters.}
\end{table}\label{Appendix_table_variables}

\subsection{Diffusion-Based Atmospheric Prior}\label{subsec0}

Atmospheric states can be represented as points in a high-dimensional phase space, where diffusion models (cite) are employed to characterize their probability distributions $P(x_0)$. The diffusion framework establishes a bijective mapping between simple distributions (e.g., Gaussian) and complex atmospheric state distributions through a forward-backward process. The forward process, defined by a Markov chain that progressively injects Gaussian noise into atmospheric states, can be formulated as a stochastic differential equation (SDE):
\begin{equation}
    \mathrm{d}\mathbf{x}=\mathbf{f}(\mathbf{x},t)\mathrm{d}t+g(t)\mathrm{d}\mathbf{w}
\end{equation}
where $x_0 \sim p_0$, $x \in \mathbb{R}^n$ represents the n-dimensional atmospheric state vector, and $w$ denotes a standard Wiener process. This SDE will gradually transform the atmospheric state into pure Gaussian noise. A remarkable result from (cite) claims that We can reverse this process and given by the reverse-time SDE:
\begin{equation}
\mathrm{d}\mathbf{x}=[\mathbf{f}(\mathbf{x},t)-g(t)^2\nabla_\mathbf{x}\log p_t(\mathbf{x})]\mathrm{d}t+g(t)\mathrm{d}\bar{\mathbf{w}}
\end{equation}
once the gradient $\nabla_\mathbf{x}\log p_t(\mathbf{x})$ (score function) is known for all $t$, we can derive this reverse SDE and simulate it to sample the high-dimensional atmospheric state  $x_0$  from simple Gaussian distributions. 

\subsection{Patch diffusion}
To model this high-dimensional atmospheric state distribution, the primary challenge lies in the inability to accommodate the uncompressed atmospheric state within GPU memory constraints (RTX4090 24GB VRAM). To address this limitation, we adopt the patch diffusion approach\cite{wang2024patch}. Rather than performing score matching on the complete atmospheric state, we propose learning the score function on randomly cropped patches from the latitude-longitude grid. The denoising score matching is conducted on these patches while preserving their corresponding geospatial positional information.

A potential limitation is that the score function, having been trained only on local patches, may inadequately capture global inter-regional dependencies. However, we justify this methodology through two key considerations. First, given that ERA5 data is generated by numerical models, which construct large-scale features from numerous local characteristics, we posit that local information sufficiently represents the majority of ERA5 dynamics. Second, we utilize relatively large patches (360×720) to ensure the model encompasses at least one complete Rossby wave cycle.

\subsection{Generative state matching}\label{subsec1}

As previously discussed, a well-trained generative model facilitates sampling from the atmospheric state distribution $P(x_0)$. This learned atmospheric prior can be utilized to modify biased forecast fields or introduce prior-consistent uncertainty into initial conditions, thereby ensuring the statistical characteristics of forecast states align with historical data. From a probabilistic perspective, this can be formulated as a Bayesian inference problem: given a prior distribution of atmospheric states $P(x_0)$ and a specific atmospheric state $x_1$, we seek to determine the posterior distribution $P(x_0|x_1)$.

To achieve this objective, we employ the SDEdit\cite{meng2021sdedit}, which utilizes SDE for high-dimensional signal manipulation. A notable advantage of the SDEdit is that it can refine high-dimensional atmospheric states without requiring additional training, provided that SDE-based generative models have been properly trained.

\begin{figure}[H]
\centering
\centerline{\includegraphics[width=\linewidth]{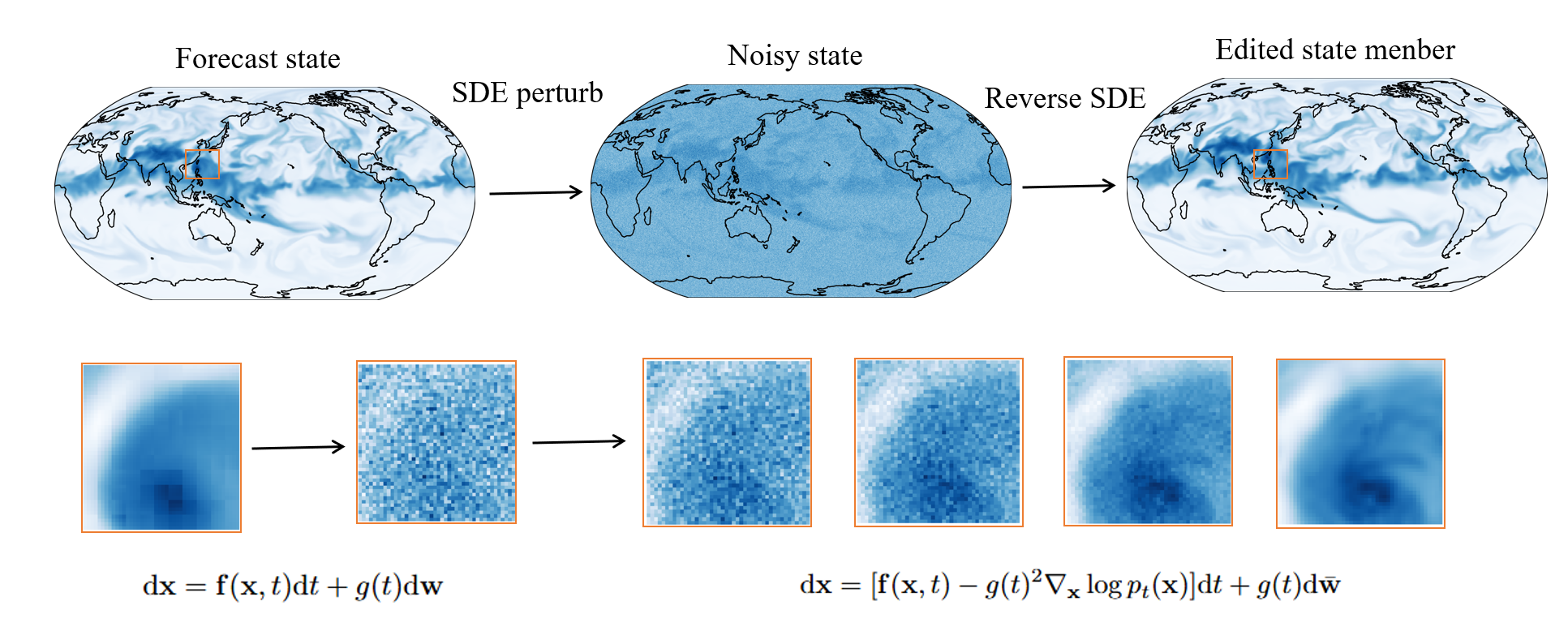}}
\caption{A conceptual interpretation of the Generalized Score Matching (GSM) framework: atmospheric states are perturbed through forward SDE, followed by trained reverse SDE that align the atmospheric state characteristics with ERA5.}\label{fig6}
\end{figure}

Figure.\ref{fig6} illustrates the implementation procedure: beginning with any atmospheric field $x_1$ (e.g., a deterministic forecast from the Pangu global weather model), SDEdit systematically perturbs the atmospheric state according to the forward diffusion process described by Equation 1. This perturbation continues until reaching a prescribed diffusion time $t_0 \in (0,T)$, where $t_0$ represents a hyperparameter (0.4 is choiced in this study). At time $t_0$, the perturbed state $x_{t_0}$ follows the conditional distribution $P(x_{t_0}|x_1)$, maintaining essential dynamical features of the original field $x_1$ while incorporating uncertainty. The process concludes by employing a reverse-time SDE (Equation 2) from $x_{t_0}$ to simulate the target posterior distribution $P(x_0|x_1)$.

Through GSM, infinite edited members can be generated from a single forecast or initial field. Iterative application of GSM enables ensemble generation for any deterministic model. Additionally, as demonstrated in Figure.\ref{fig6}, GSM decouples the alignment of atmospheric states with objective distributions from model-specific systematic biases, facilitating interactive integration across different models for cross-model ensemble implementation.

\subsection{Ensemble of Expert Denoisers}

Given the computational complexity of learning high-dimensional atmospheric distributions, while a highly expressive neural network is essential for this task, hardware constraints (particularly GPU memory limitations) preclude simple scaling of network capacity through increased feature channels or layer depth.

Drawing inspiration from ensemble diffusion approaches\cite{balaji2022ediff}, we enhance GSM performance through a two-stage training strategy:  first training a comprehensive diffusion model across all $t \in (0,T)$, followed by stage-specific fine-tuning for a designated $(t_0,T)$ . The ensemble diffusion framework posits that posterior inference tasks vary across different temporal stages: initial denoising, such as is predominantly governed by prior distributions, while later stages  are more condition-dependent. This necessitates specific fine-tuning for t0, with refinement concentrated on the $(t_0,T)$ interval for GSM. This approach yields substantial performance improvements while maintaining inference computational costs.

\subsection{Diffusion model specification}

\subsubsection{Score estimation formulation}
We adopt the Denoising Diffusion Probabilistic Models (DDPM) \cite{ho2020denoising} formulation, discretizing the forward process into 1000 steps. The noise schedule follows a linear progression from $\beta_1 = 10^{-4}$ to $\beta_T = 0.02$. The forward process can be expressed as:
\begin{equation}
q(\mathbf{x}_t|\mathbf{x}_0) = \mathcal{N}(\mathbf{x}_t; \sqrt{\alpha_t}\mathbf{x}_0, (1-\alpha_t)\mathbf{I})
\label{eq:forward}
\end{equation}
where $\alpha_t := \prod_{s=1}^t \alpha_s$ and $\alpha_t := 1 - \beta_t$.
A model can be trained to predict the added noise using a simple loss function \cite{ho2020denoising}:
\begin{equation}
L_{simple}(\theta) := \mathbb{E}_{t,\mathbf{x}0,\epsilon}\left[\left|\epsilon - \epsilon_\theta(\sqrt{\alpha_t}\mathbf{x}_0 + \sqrt{1-\alpha_t}\epsilon, t)\right|^2\right]
\label{eq:loss_SI}
\end{equation}
Through Tweedie's formula, we obtain:
\begin{equation}
\nabla \log p(\mathbf{x}_t) = -\frac{1}{1-\alpha_t}\epsilon
\label{eq:tweedie}
\end{equation}
demonstrating that noise estimation is equivalent to estimating a scaled version of the score function. This enables estimation of the score function across all timesteps $t$.

\subsubsection{Backbone and optimisation}

This study employs U-ConvNext as the backbone architecture for the diffusion model. The network architecture comprises three distinct stages: encoder, middle, and decoder stages. The input data undergoes downsampling after the encoder stage before entering the middle stage, followed by upsampling into the decoder stage, with skip connections implemented prior to upsampling. The encoder and decoder stages each contain 12 ConvNext blocks\cite{liu2022convnet}, while the middle stage incorporates 24 ConvNext blocks. The feature dimensions are set to 256 and 512 for the respective stages.

The model training was conducted using 6 NVIDIA RTX 4090 GPUs, with a batch size of 1 per GPU, yielding a total batch size of 6. Optimization was performed using the Adam optimizer with a constant learning rate of $2.5 \times 10^{-4}$. The training of the complete unconditional diffusion model spanned three weeks, followed by an additional week of fine-tuning focused on the diffusion process for $t \in [0.6, 1]$.

\subsection{Evaluation method}\label{Appendix_method_B}

We evaluated various aspects of GenEPS forecast skill using multiple verification metrics: the Anomaly Correlation Coefficient (ACC), Root Mean Square Error (RMSE), Continuous Ranked Probability Score (CRPS), and ensemble SPREAD. Each metric assesses distinct characteristics of the forecasting system's performance.

The ACC quantifies the spatial correlation between ensemble mean forecast anomalies and ERA5 anomalies relative to climatology mean. For a weather variable $\nu$ at time $t$, ACC is defined as:

\begin{equation}
\text{ACC}(\nu, t) = \frac{\sum_{i,j} L(i) (\hat{A}_{i,j,t}^\nu-M_{i,j,t}^\nu) (A_{i,j,t}^\nu-M_{i,j,t}^\nu)}{\sqrt{\sum_{i,j} L(i) \left( \hat{A}_{i,j,t}^\nu-M_{i,j,t}^\nu \right)^2 \times \sum_{i,j} L(i) \left( A_{i,j,t}^\nu-M_{i,j,t}^\nu \right)^2}}
\end{equation}
for ACC, closer to 1 is better.

where $\hat{A}_{i,j,t}^\nu$ represents the ensemble mean forecast at coordinate $(i,j)$, and $L(i) = N_{\text{lat}} \times \frac{\cos \phi_i}{\sum_{i=1}^{N_{\text{lat}}} \cos \phi_i}$ is the latitudinal weight at latitude $\phi_i$. $A_{i,j,t}^\nu$ denotes the ERA5 value,  $M_{i,j,t}^\nu$ is the climatology mean, which is computed using 39-year historic data. The metric can be evaluated globally or for specific regions (Northern Hemispheres).

Using the same notations as ACC, RMSE can be defined as:


\begin{equation}
\text{RMSE}(\nu, t) = \sqrt{\frac{\sum_{i=1}^{N_{\text{lat}}} \sum_{j=1}^{N_{\text{lon}}} L(i) (\hat{A}_{i,j,t}^\nu - A_{i,j,t}^\nu)^2}{{N_{\text{lat}}} \times {N_{\text{lon}}}}}
\end{equation}

for RMSE, smaller is better.

CRPS is used to quantifies the discrepancy between the ensemble forecasted cumulative distribution function (CDF) and the ERA5 CDF, defined as, 

\begin{equation}
\text{CRPS} = L(i)\int_{-\infty}^{+\infty} \left[ F(\hat{A}^{u}_{i,j,t}) - \mathbb{I}(\hat{A}^{u}_{i,j,t} \leq z) \right] dz
\end{equation}

where \( F(\cdot) \) denotes the cumulative distribution function of the forecast distribution and \( \mathbb{I}(\cdot) \) is an indicator function that takes a value of 1 if the statement is true and 0 otherwise. In deterministic forecasts, the CRPS is equal to the mean
absolute error (MAE).

Spread quantifies the forecast uncertainty of the ensemble members, defined as:
\begin{equation}
\text{SPREAD}(\nu, t) = \sqrt{\frac{\sum_{i=1}^{N_{\text{lat}}} \sum_{j=1}^{N_{\text{lon}}} L(i) \text var(\hat{A}_{i,j,t}^\nu)}{{N_{\text{lat}}} \times {N_{\text{lon}}}}}
\end{equation}

\backmatter




\bibliography{sn-bibliography}

\appendix


\end{document}